\documentclass{rspublic-date}
\usepackage{graphicx}

\newcommand{\del}{\mbox{\boldmath{$\nabla$}}}
\newcommand{\nhat}{\mbox{\boldmath{$\hat{n}$}}}

\newcommand{\xhat}{\mbox{\boldmath{$\hat{x}$}}}
\newcommand{\yhat}{\mbox{\boldmath{$\hat{y}$}}}
\newcommand{\phib}{\mbox{\boldmath{$\phi$}}}
\newcommand{\Jb}{\mbox{{\bf J}}}
\newcommand{\Fb}{\mbox{{\bf F}}}
\newcommand{\Eb}{\mbox{{\bf E}}}
\newcommand{\ub}{\mbox{{\bf u}}}
\newcommand{\wb}{\mbox{{\bf w}}}
\newcommand{\fb}{\mbox{{\bf f}}}
\newcommand{\erf}{\mbox{erf\hspace{2pt}}}
\newcommand{\Pe}{\mbox{P\hspace{-1pt}e\hspace{2pt}}}
\newcommand{\Nu}{\mbox{N\hspace{-1pt}u\hspace{2pt}}}

\begin{document}

%
%
%
%

\title[Conformal mapping of some non-harmonic functions] { Conformal
mapping of \\ some non-harmonic functions\\ in transport theory }
\author[M. Z. Bazant]{Martin Z. Bazant}
\affiliation{ Department of Mathematics\\ Massachusetts Institute of
Technology, Cambridge, MA 02139 USA\\ {\em and}\\
Laboratoire de Physico-chimie Th\'eorique\\
\'Ecole
Sup\'erieure de Physique et de Chimie Industrielles\\ 10 rue Vauquelin,
75231 Paris, France.}

\date{\today}
\label{firstpage}
\maketitle

\begin{abstract}{conformal mapping, non-harmonic functions, nonlinear
diffusion, advection-diffusion, electrochemical transport}

Conformal mapping has been applied mostly to harmonic functions, i.e.
solutions of Laplace's equation. In this paper, it is noted that some
other equations are also conformally invariant and thus equally well
suited for conformal mapping in two dimensions. In physics, these
include steady states of various nonlinear diffusion equations, the
advection-diffusion equations for potential flows, and the
Nernst-Planck equations for bulk electrochemical transport.  Exact
solutions for complicated geometries are obtained by conformal mapping
to simple geometries in the usual way. Novel examples include
nonlinear advection-diffusion layers around absorbing objects and
concentration polarizations in electrochemical cells. Although some of
these results could be obtained by other methods, such as Boussinesq's
streamline coordinates, the present approach is based on a simple
unifying principle of more general applicability.  It reveals a basic
geometrical equivalence of similarity solutions for a broad class of
transport processes and paves the way for new applications of
conformal mapping, e.g.  to non-Laplacian fractal growth.

\end{abstract}

\section{ Introduction }

Complex analysis is one of the most beautiful subjects in mathematics, and,
in spite of involving imaginary numbers, it has remarkable relevance for
`real' applications. One of its most useful techniques is conformal
mapping, which transforms planar domains according to analytic functions,
$w = f(z)$, with $f^\prime(z) \neq 0$.  Geometrically, such mappings induce
upon the plane a uniform, local stretching by $|f^\prime(z)|$ and a 
rotation by $\mbox{arg} \ f^\prime(z)$.  This `ampli-twist' interpretation
of the derivative implies conformality, the preservation of angles between
intersecting curves (Needham 1997).

The classical application of conformal mapping is to
solve Laplace's equation,
\begin{equation}
\del^2\phi=0 ,  \label{eq:laplace}
\end{equation}
i.e. to determine harmonic functions, in complicated planar domains by
mapping to simple domains. The method relies on the conformal invariance of
Eq.~(\ref{eq:laplace}), which remains the same after a conformal change of
variables.
Before the advent of computers, important analytical solutions were thus
obtained for electric fields in capacitors, thermal fluxes around pipes,
inviscid flows past airfoils, etc. (Needham 1997; Churchill \& Brown 1990;
Batchelor 1967).  Today, conformal mapping is still used extensively in
numerical methods (Trefethen 1986).

Currently in physics, a veritable renaissance in conformal mapping is
centering around `Laplacian-growth' phenomena, in which the motion of a
free boundary is determined by the normal derivative of a harmonic
function. Continuous problems of this type include viscous fingering, where
the pressure is harmonic (Saffman \& Taylor 1958; Bensimon {\it et al.}
1986; Saffman 1986),
and solidification from a supercooled melt, where the temperature is
harmonic in some approximations (Kessler {\it et al.} 1988; Cummings {\it
et al.} 1999). Such problems can be elegantly formulated in terms of
time-dependent conformal maps, which generate the moving boundary from its
initial position. This idea was first developed by Polubarinova-Kochina
(1945a; 1945b) and Galin (1945) with recent interest stimulated by Shraiman
\& Bensimon (1984) focusing on finite-time singularities and pattern
selection (Howison 1986; Tanveer 1987; Dai {\it et al.} 1991; Ben Amar
1991; Howison 1992; Tanveer 1993; Ben Amar \& Brener 1996; Ben Amar \&
Poir\'e 1999; Feigenbaum {\it et al.} 2001).

Stochastic problems of a similar type include diffusion-limited aggregation
(DLA) (Witten \& Sander 1981) and dielectric breakdown (Niemeyer {\it et
al.} 1984). Recently, Hastings \& Levitov (1998) proposed an analogous
method to describe DLA using iterated conformal maps, which initiated a
flurry of activity applying conformal mapping to Laplacian fractal-growth
phenomena (Davidovitch {\it et al.} 1999, 2000; Barra {\it et al.} 2002a,
2002b; Stepanov \& Levitov 2001; Hastings 2001; Somfai {\it et al.}  1999;
Ball \& Somfai 2002). One of our motivations here is to extend such
powerful analytical methods to fractal growth phenomena limited by {\it
non-Laplacian} transport processes.

Compared to the vast literature on conformal mapping for Laplace's
equation, the technique has scarcely been applied to any other
equations. The difficulty with non-harmonic functions is illustrated
by Helmholtz's equation,
\begin{equation}
\del^2 \phi = \phi,  \label{eq:helm}
\end{equation}
which arises in transient diffusion and electromagnetic radiation (Morse \&
Feshbach 1953). After conformal mapping, $w = f(z)$, it acquires a
cumbersome, non-constant coefficient (the Jacobian of the map),
\begin{equation}
|f^\prime |^2 \, \del^2 \phi = \phi . \label{eq:helmtrans}
\end{equation}
Similarly, the bi-harmonic equation,
\begin{equation}
\del^2 \del^2 \phi = 0 , \label{eq:bi}
\end{equation}
which arises in two-dimensional viscous flows (Batchelor 1967) and
elasticity (Muskhelishvili 1953), transforms with an extra Laplacian
term (see below),
\begin{equation}
|f^\prime|^4 \del^2 \del^2 \phi = - 4\, |f^{\prime\prime}|^2 \del^2\phi
 . \label{eq:bitrans}
\end{equation}
In this special case, conformal mapping is commonly used (e.g. Chan {\it et
al.} 1997; Crowdy 1999, 2002; Barra {\it et al.} 2002b) because solutions
can be expressed in terms of analytic functions in Goursat form
(Muskhelishvili 1953). Nevertheless, given the singular ease of applying
conformal mapping to Laplace's equation, it is natural to ask whether any
other equations share its conformal invariance, which is widely believed to
be unique.

In this paper, we show that certain {\it systems} of nonlinear
equations, with non-harmonic solutions, are also conformally
invariant. In section~\ref{sec:math}, we give a simple proof of this
fact and some of its consequences. In section~\ref{sec:phys}, we
discuss applications to nonlinear diffusion phenomena and show that
single conformally invariant equations can always be reduced to
Laplace's equation (which is not true for coupled systems). In
section~\ref{sec:ad}, we apply conformal mapping to nonlinear
advection-diffusion in a potential flow, which is equivalent to
streamline coordinates in a special case (Boussinesq 1905). In
section~\ref{sec:elec}, we apply conformal mapping to nonlinear
electrochemical transport, apparently for the first time. In
section~\ref{sec:concl}, we summarize the main results. Applications
to non-Laplacian fractal growth are a common theme throughout the
paper. (See sections 2, 4, and 6.)

\section{ Mathematical Theory }
\label{sec:math}

\subsection{ Conformal Mapping without Laplace's Equation?  } 
\begin{figure}
\begin{center}
\includegraphics[width=3.5in]{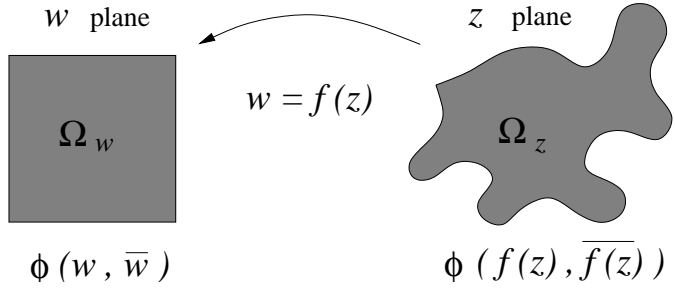}
\caption{ Conformal mapping, $w = f(z)$, of a solution, $\phi$, to a
conformally invariant equation from a complicated domain, $\Omega_z$,
and a simple domain, $\Omega_w$. \label{fig:cartoon} }
\end{center}
\end{figure}

The
standard application of conformal mapping is based on two facts:
\begin{enumerate}
\item Any harmonic function, $\phi$, in a singly connected planar
domain, $\Omega_w$, is the real part of a analytic function, $\Phi$, the
`complex potential' (which is unique up to an additive constant): $\phi =
\Real \, \Phi(w)$.
\item Since analyticity is preserved under composition, harmonicity is
preserved under conformal mapping, so $\phi = \Real \,
\Phi(f(z))$ is harmonic in $\Omega_z = f^{-1}(\Omega_w)$.
\end{enumerate}
Presented like this, it seems that conformal mapping is closely tied to
harmonic functions, but Fact 2 simply expresses the conformal invariance of
Laplace's equation: A solution, $\phi(w)$, is the same in any mapped
coordinate system, $\phi(f(z))$. Fact 1, a special relation between
harmonic functions and analytic functions, is not really needed. If another
equation were also conformally invariant, then its non-harmonic solutions,
$\phi(w,\overline{w})$, would be preserved under conformal mapping in the
same way, $\phi(f(z),\overline{f(z)})$. (See Fig.~\ref{fig:cartoon}.)

In order to seek such non-Laplacian invariant equations, we review the
transformation properties of some basic differential operators. Following
Argand and Gauss, it is convenient to represent two-dimensional vectors,
${\bf a} = a_x \xhat + a_y \yhat$, as complex numbers, $a = a_x + a_y i$.
We thus express the gradient vector operator in the plane as a complex
scalar operator\footnote{ Although $\frac{\partial}{\partial \overline{z}}
= \frac{1}{2}\nabla$ is more common in the mathematical literature, we
prefer $\nabla$ for applications in transport theory because gradients play
a central role.},
\begin{equation}
\del = \xhat \frac{\partial }{\partial x} + \yhat \frac{\partial }{\partial
y}  \ \ \ \longleftrightarrow \ \ \ \nabla = \frac{\partial }{\partial x} +
i \frac{\partial }{\partial y} ,
\end{equation}
which has the essential property that $\nabla f = 0$ if and only if $f$ is
analytic, in which case, $\overline{\nabla} f = 2 f^\prime$ (Needham 1997).
Since ${\bf a}\cdot{\bf b} = \Real \, a\overline{b}$, the Laplacian
operator can be expressed as, $\del \cdot \del = \Real \, \nabla
\overline{\nabla} = \nabla \overline{\nabla}$ (if mixed partial derivatives
can be taken in any order). Similarly, the `advection operator', which acts
on {\it two} real functions $\phi$ and $c$, takes the form, $\del \phi
\cdot \del c = \Real \, (\nabla \phi)\overline{\nabla}c$.

Under a conformal mapping of the plane, $w=f(z)$, the gradient transforms
as,
$\nabla_z = \overline{f^\prime}\, \nabla_w$.
This basic fact, combining the ampli-twist property and the chain rule,
makes it easy to transform differential operators (Needham 1997). The
Laplacian transforms as,
\begin{equation}
\nabla_z \overline{\nabla}_z  
= (\nabla_z f^\prime)\overline{\nabla}_w + |f^\prime|^2 \nabla_w
\overline{\nabla}_w =  |f^\prime|^2 \, \nabla_w  \overline{\nabla}_w
\label{eq:lapl}
\end{equation}
where $\nabla_z f^\prime =0$ because $f^\prime$ is also analytic.
This immediately implies the conformal invariance of Laplace's equation
(\ref{eq:laplace}), and the non-invariance of Helmholtz's equation
(\ref{eq:helm}). The transformation of the bi-harmonic equation
(\ref{eq:bi}) in Eq.~(\ref{eq:bitrans}) is also easily derived with the
help of Needham's identity, $\Delta |f|^2 = 4 |f^\prime|^2$, applied to
$f^\prime$.

Everything in this paper follows from the simple observation
that the advection operator transforms just like the Laplacian,
\begin{equation}
\Real \, (\nabla_z \phi)\, \overline{\nabla}_z c = |f^\prime|^2 \, \Real \,
(\nabla_w \phi)\, \overline{\nabla}_w c . \label{eq:conv}
\end{equation}
Each operator involves a `dot product of two gradients', so the same
Jacobian factor, $|f^\prime|^2$, appears in both cases. The
transformation laws, Eq.~(\ref{eq:lapl}) and Eq.~(\ref{eq:conv}), are
surely well known, but it seems that some general implications have
been overlooked, or at least not fully exploited in physical
applications.

\subsection{ Conformally Invariant Systems of Equations }

The identities ~(\ref{eq:lapl}) and (\ref{eq:conv}) imply the conformal
invariance of any  system of equations of the general form,
\begin{equation}
\sum_{i=1}^N \left( \, a_i(\phib) \,\del^2 \phi_i + \sum_{j=i}^N
\,a_{ij}(\phib) \,\del \phi_i \cdot \del \phi_j\, \right) = 0
\label{eq:geneq}
\end{equation}
where the coefficients $a_i(\phib)$ and $a_{ij}(\phib)$ may be nonlinear
functions of the unknowns, $\phib = (\phi_1, \phi_2, \ldots, \phi_N)$, but
not of the independent variables or any derivatives of the unknowns. Thus
we arrive at our main result:

\begin{theorem}{\em (Conformal Mapping Theorem.)}  Let
$\phib\left(w,\overline{w}\right)$  satisfy Eq.~(\ref{eq:geneq}) in a
domain $\Omega_w$ of the complex plane, and let $w=f(z)$ be a conformal
mapping from $\Omega_z$ to $\Omega_w$. Then $\phib(f(z),\overline{f(z)})$
satisfies Eq.~(\ref{eq:geneq}) in $\Omega_z$.
\end{theorem}

Whenever the system (\ref{eq:geneq}) can be solved analytically in some
simple domain, the Theorem produces a family of exact solutions
for all topologically equivalent domains. Otherwise, it allows a
convenient numerical solution to be mapped to more  complicated
domains of interest. This is an enormous simplification
for free boundary problems, where the  solution in an evolving domain can be
obtained by time-dependent conformal mapping to a single, static domain.


Conformal mapping is most useful when the boundary
conditions are also invariant. Dirichlet ($\phi_i=$ constant) or Neumann
($\nhat\cdot\del\phi_i=0$) conditions are typically assumed, but here we
consider the straight-forward generalizations,
\begin{equation}
b_{i}(\phib)=0 \ \ \ \mbox{ and } \ \ \  \sum_{j=1}^N\, b_{ij}(\phib) \,
(\nhat\cdot\del\phi_j)^{\alpha_i} = 0 \label{eq:genbc}
\end{equation}
respectively, where $b_{i}(\phib)$ and $b_{ij}(\phib)$ are nonlinear
functions of the unknowns, $\alpha_i$ is a constant, and $\nhat$ is
the unit normal. The conformal invariance of the former is obvious, so
we briefly consider the latter.

It is convenient to locally transform a vector field, $F$, along a given
contour as, $\tilde{F} = t\, \overline{F}$, so that $\Real\, \tilde{F}$ and
$\Imag \, \tilde{F}$ are the projections onto the unit tangent,
$t=dz/|dz|$, and the (right-handed) unit normal, $n = -i t$, respectively.
Since the tangent transforms as, $t_w = t_z f^\prime/|f^\prime|$, and the
gradient as, $\nabla_z = \overline{f^\prime} \nabla_w$, we find,
$\tilde{\nabla}_z = |f^\prime| \tilde{\nabla}_w$. The invariance of
Eq.~(\ref{eq:genbc}) follows after taking the imaginary part on the
boundary contour.

\subsection{ Gradient-Driven Flux Densities }  Generalizing $\del
\phi$ for Laplacian problems, we define a `flux density' for solutions of
Eq.~(\ref{eq:geneq}) to be any quasi-linear combination of gradients,
\begin{equation}
\Fb_i
= \sum_{j=1}^N c_{ij}(\phib) \del \phi_j , \label{eq:fluxdef}
\end{equation}
where $c_{ij}(\phib)$ are nonlinear functions. The transformation rules
above for the gradient apply more generally to any flux density,
\begin{equation}
F_z = \overline{f^\prime}\, F_w \ \ \ \mbox{and} \ \ \ \tilde{F}_z =
|f^\prime|\, \tilde{F}_w .  \label{eq:Ftrans}
\end{equation}
These basic identities imply a curious geometrical equivalence between
solutions to {\it different} conformally invariant systems:

\begin{theorem}{\em (Equivalence Theorem.)} Let $\phib^{(1)}$ and
$\phib^{(2)}$ satisfy equations of the form (\ref{eq:geneq}) with
corresponding flux densities, $F^{(1)}$ and $F^{(2)}$, of the form
(\ref{eq:fluxdef}). If $F^{(1)}_z = a \, F^{(2)}_z$ on a contour $C_z$ for
some complex constant $a$, then $F^{(1)}_w = a \, F^{(2)}_w$ on the image,
$C_w = f(C_z)$, after a conformal  mapping, $w = f(z)$.
\end{theorem}

\noindent An important corollary pertains to `similarity solutions' of
Eqs.~(\ref{eq:geneq}) and (\ref{eq:genbc}) in which certain variables
$\{\phi_i\}$ involved in a flux density depend on only one Cartesian
coordinate, say $\Real w$, after conformal mapping: $\phi_i =
G_i(\Real f(z))$.  (Our examples below are mostly of this type.)  Such
special solutions share the same flux lines (level curves of $\Imag
f(z)$) and iso-potentials (level curves of $\Real f(z)$) in any
geometry attainable by conformal mapping. They also share the same
{\it spatial distribution} of flux density on an iso-potential,
although the {\it magnitudes} generally differ.


An important physical quantity is the total normal flux through a
contour, often called the `Nusselt number', $\Nu$. For any contour,
$C$, we define a complex total flux, $I(C) = \int_C \tilde{F}\, |dz| =
\int_C \overline{F}\, dz$, \label{eq:I} such that $\Real I(C)$ is the
integrated tangential flux and $\Imag I(C) = \Nu(C)$. From
Eq.~(\ref{eq:Ftrans}) and $dw = f^\prime dz$, we conclude, $I(C_z) =
I(C_w)$. Therefore, flux integrals can be calculated in any convenient
geometry.

This basic fact has many applications. For example, if $\tilde{F}_w$
is constant on a contour $C_w = f(C_z)$, then for any conformal
mapping, we have, $I(C_z) = I(C_w) = \ell(C_w) \tilde{F}_w$, where
$\ell(C_w)$ is simply the length of $C_w$. For fluxes driven by
gradients of harmonic functions, this is the basis for the method of
iterated conformal maps for DLA, in which the `harmonic measure' for
random growth events on a fractal cluster is replaced by a uniform
probability measure on the unit circle (Hastings \& Levitov
1998). 

More generally, a non-harmonic probability measure for fractal growth
can be constructed for any flux law of the form (\ref{eq:fluxdef}) for
fields satisfying Eq.~(\ref{eq:geneq}). According to the results
above, the growth probability is simply proportional to the normal
flux density on the unit circle for the same transport problem after
conformal mapping to the exterior of the unit disk. A nontrivial
example is given below in section~\ref{sec:ad}(c). This allows the
Hastings-Levitov method to be extended to a broad class of
non-Laplacian fractal-growth processes (Bazant, Choi \& Davidovitch
2003).

\subsection{ Conformal Mapping to Curved Surfaces }

The Conformal Mapping Theorem is even more general than it might
appear from our proof: The domain, $\Omega_z$, may be contained in any
two-dimensional manifold. This becomes clear from the recent work of
Entov and Etingof (1991; 1997), who solved viscous fingering problems
on various curved surfaces by conformal mapping to the complex plane,
e.g. via stereographic projection from the Riemann sphere. They
exploited the fact that Laplace's equation is invariant under any
conformal mapping, $\wb = \fb(z)$, from the plane to a curved surface
because the Laplacian transforms as $\del_z^2 = J \, \del^2_w$, where
$J(\fb(z))$ is the Jacobian. The system (\ref{eq:geneq}) shares this
general conformal invariance because the advection operator transforms
in the same way, $\del_z \phi \cdot \del_z c = J \, \del_w \phi \cdot
\del_w c$. The application of these ideas to non-Laplacian
transport-limited growth phenomena on curved surfaces is work in
progress with J. Choi and D. Crowdy; here we focus on conformal
mappings in the plane, described by analytic functions.

\section{ Physical Applications to Diffusion Phenomena }
\label{sec:phys}

Conformally invariant boundary-value problems of the form (\ref{eq:geneq})
and (\ref{eq:genbc}) commonly arise in physics from steady conservation
laws,
\begin{equation}
\frac{\partial c_i}{\partial t} = \del\cdot\Fb_i = 0 ,  \label{eq:cons}
\end{equation}
for gradient-driven flux densities, Eq.~(\ref{eq:fluxdef}), with algebraic
($b(c_i)=0$) or zero-flux ($\nhat\cdot\Fb_i=0$) boundary conditions, where
$c_i$ is the concentration and $\Fb_i$ the flux of substance $i$.
Hereafter, we focus on flux densities of the form,
\begin{equation}
\Fb_i = c_i\, \ub_i - D_i(c_i)\, \del c_i, \ \ \  \ub_i \propto
\del \phi  \label{eq:Fi}
\end{equation}
where $D_i(c_i)$ is a nonlinear diffusivity, $\ub_i$ is an
irrotational vector field causing advection, and $\phi$ is a (possibly
non-harmonic) potential. Examples include s advection-diffusion in
potential flows and bulk electrochemical transport.

Before discussing these cases of coupled dependent variables, it is
instructive to consider nonlinear diffusion in only one variable. The most
general equation of the type (\ref{eq:geneq}) for one variable is,
\begin{equation}
a(c)\, \del^2 c = |\del c|^2 .  \label{eq:ivan}
\end{equation}
This equation arises in the Stefan problem of dendritic solidification,
where $c$ is the dimensionless temperature of a supercooled melt and $a(c)$
is Ivantsov's function, which implicitly determines the position of the
liquid-solid interface via $a(c)=1$ (Ivantsov 1947). In two dimensions,
Bedia \& Ben Amar (1994) prove the conformal invariance of
Eq.~(\ref{eq:ivan}) and then study similarity solutions, $c(\xi,\eta) =
G(\eta)$, by conformal mapping, $w=\xi + i\eta$, to a plane of parallel
flux lines,
\begin{equation}
a(G)\, G^{\prime\prime} = (G^\prime)^2 \label{eq:G} ,
\end{equation}
where an ordinary differential equation is solved.

More generally, reversing these steps, it is straight-forward to show that
any monotonic solution of Eq.~(\ref{eq:G}) produces a nonlinear
transformation, $c = G(\phi)$, from Eq.~(\ref{eq:ivan}) to Laplace's
equation (\ref{eq:laplace}), which implies conformal invariance. There are
several famous examples. For steady concentration-dependent diffusion,
\begin{equation}
\del \cdot ( D(c) \del c) = 0 ,
\end{equation}
it is Kirchhoff's transformation 
(Crank 1975),
 $\phi = G^{-1}(c) = \int_0^c D(x) dx$.
For Burgers' equation in an irrotational flow ($\ub = -\nabla h$),
\begin{equation}
\frac{\partial \ub}{\partial t} + \lambda \ub \cdot \del \ub = \nu \del^2
\ub ,
\end{equation}
which is equivalent to the KPZ
equation without noise (Kardar {\it et al.} 1986),
\begin{equation}
\frac{\partial h}{\partial t} = \nu \del^2 h + \frac{\lambda}{2} |\del
h|^2 ,
  \label{eq:KPZ}
\end{equation}
it is the Cole-Hopf transformation  (Whitham 1974),
 $\phi = G^{-1}(h) = e^{\lambda h/2\nu}$,
which yields the diffusion equation,
 $\frac{\partial \phi}{\partial t} = \nu \del^2 \phi$,
and thus Laplace's equation in steady state.

In summary, the general solutions to Equation~(\ref{eq:ivan}) are simply
nonlinear functions of harmonic functions, so, in the case of one variable,
our theorems can be easily understood in terms of standard conformal
mapping. For two or more coupled variables, however, this is no longer
true, except for special similarity solutions.  The following sections
discuss some truly non-Laplacian physical problems.

\section{ Steady Advection-Diffusion in a Potential Flow}
\label{sec:ad}

We begin with a well known system of the form (\ref{eq:genbc}), the only
one to which conformal mapping has previously been applied (see below),
albeit not in the present, more general context. Consider the steady
diffusion of particles or heat passively advected in a potential flow,
allowing for a concentration-dependent diffusivity. For a characteristic
length, $L$, speed, $U$, concentration, $C$, and diffusivity, $D(C)$, the
dimensionless equations are
\begin{equation}
\del^2 \phi = 0 \ \ \ \mbox{and} \ \ \ \Pe \, \del\phi \cdot \del c =
\del\cdot( b(c)\,  \del c) ,\label{eq:cd}
\end{equation}
where $\phi$ is the velocity potential (scaled to $U L$), $c$ is the
concentration (scaled to $C$), $b(c)$ is the dimensionless diffusivity, and
$\Pe = UL/D$ is the P\'eclet number. The latter equation is a steady
conservation law for the dimensionless flux density,
$\Fb =  \Pe \, c \del \phi\,  -\, b(c)\,  \del c$
(scaled to $D C/L$). For $b(c)=1$, these classical equations have been
studied recently in two dimensions, e.g. in the contexts of tracer
dispersion in porous media (Koplik {\it et al.} 1994, 1995), vorticity
diffusion in strained wakes (Eames \& Bush 1999; Hunt \& Eames 2002),
thermal advection-diffusion (Morega \& Behan 1994; Sen \& Yang 2000), and
dendritic solidification in flowing melts (Kornev \& Mukhamadullina 1994;
Cummings {\it et al.} 1999).

\subsection{ Similarity Solutions for Absorbing Leading Edges }
\begin{figure}
\includegraphics[width=2.4in]{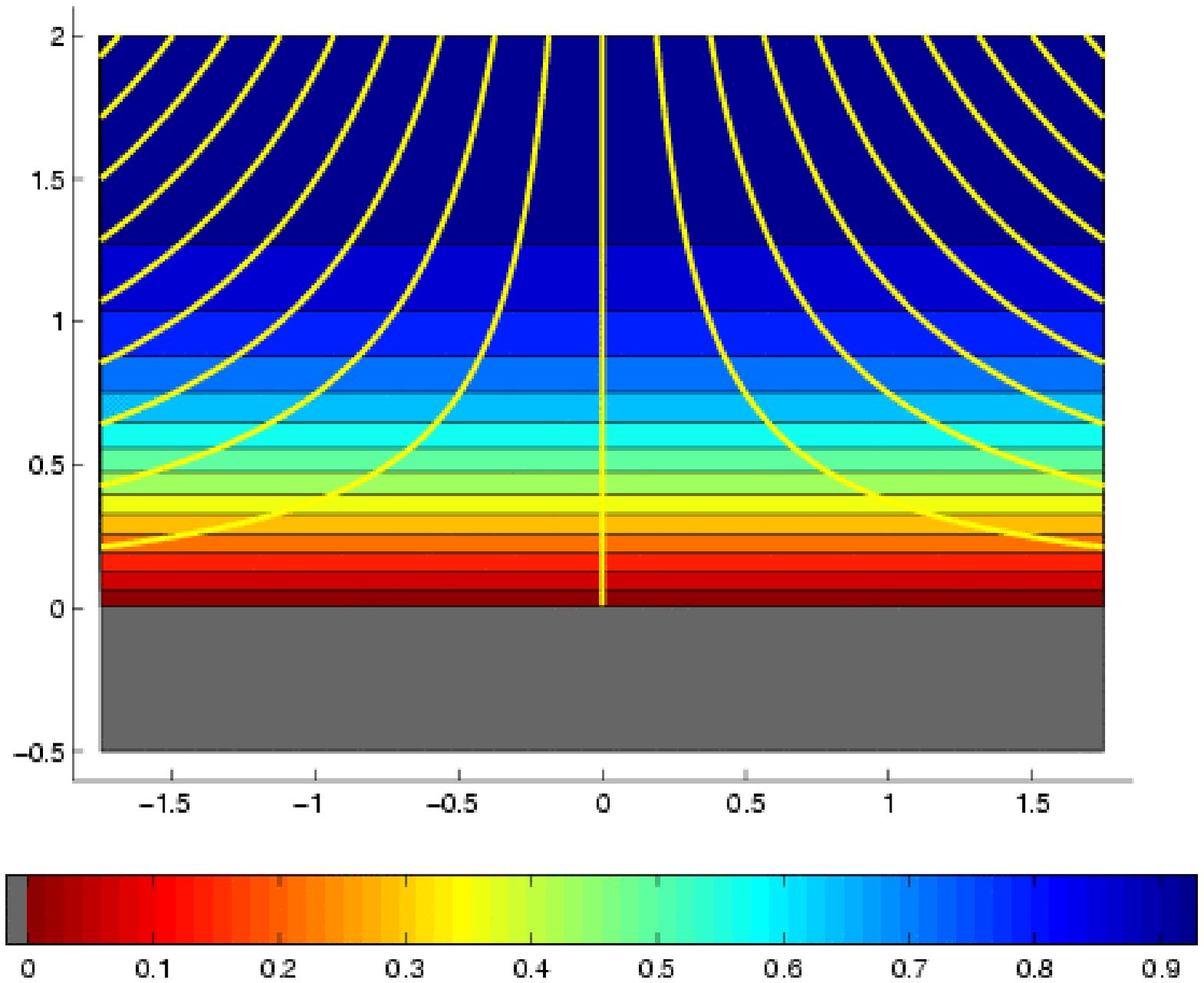} \nolinebreak
\includegraphics[width=2.4in]{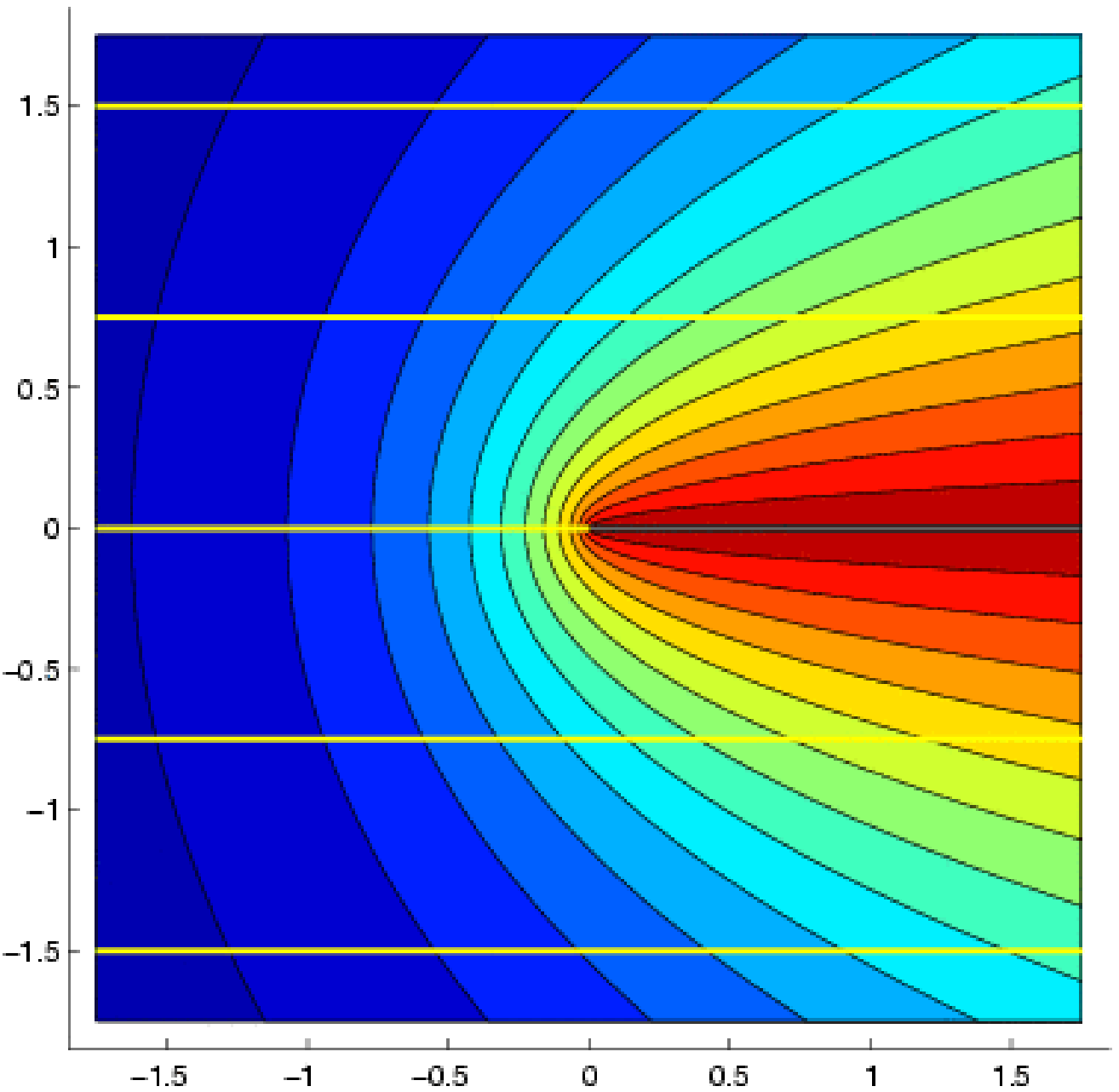} \\
\includegraphics[width=2.4in]{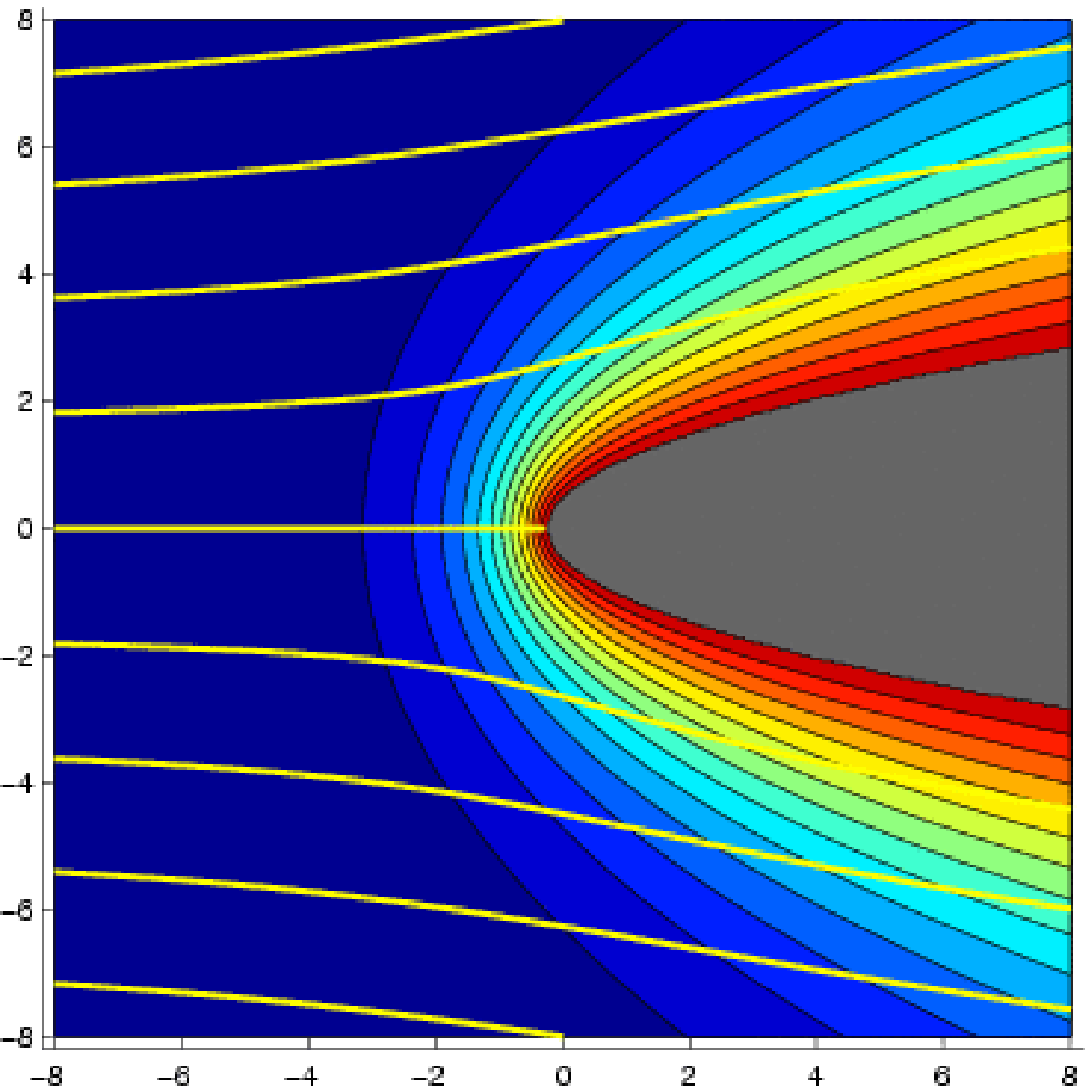} \nolinebreak
\includegraphics[width=2.4in]{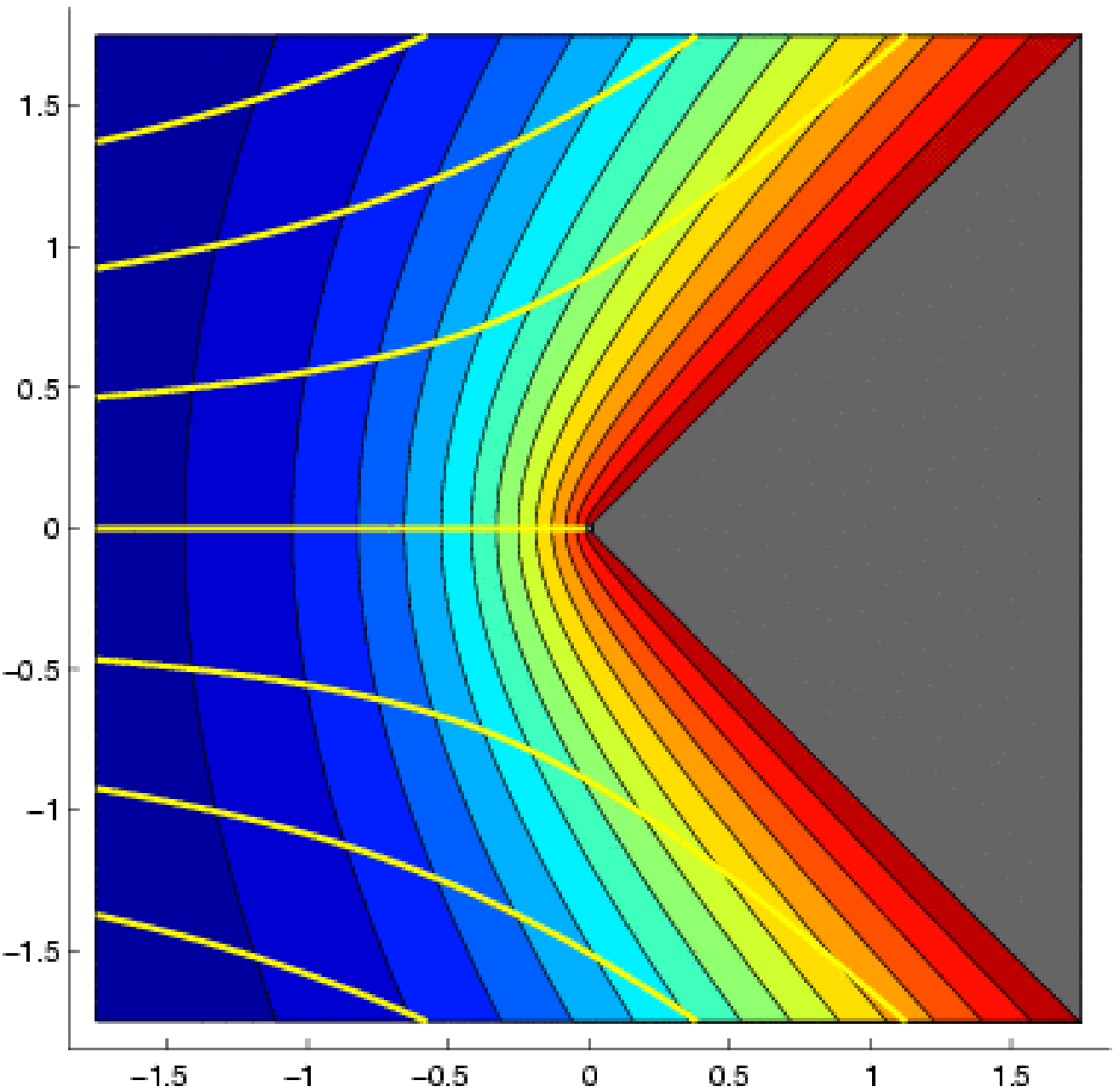} \\
\includegraphics[width=2.4in]{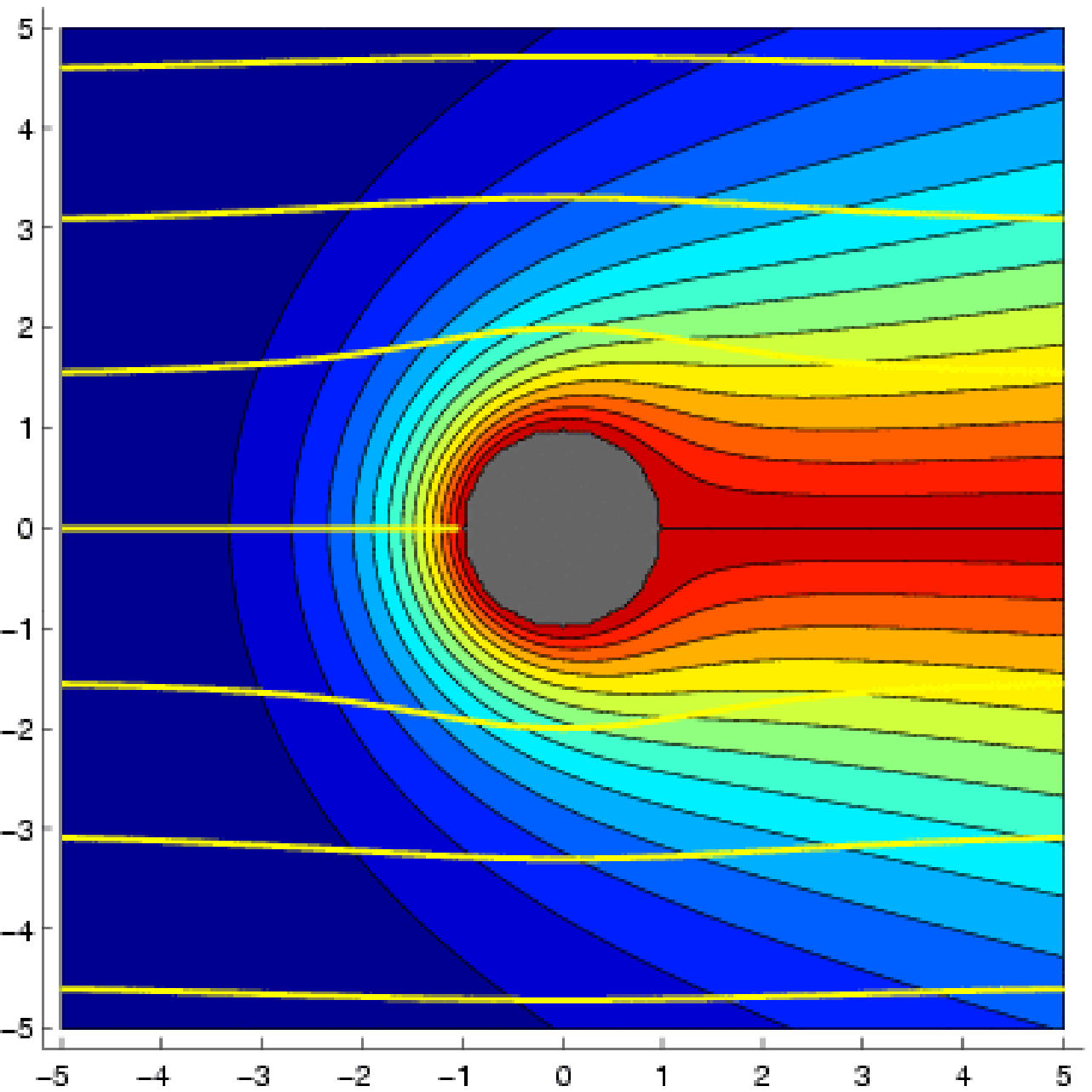} \nolinebreak
\includegraphics[width=2.4in]{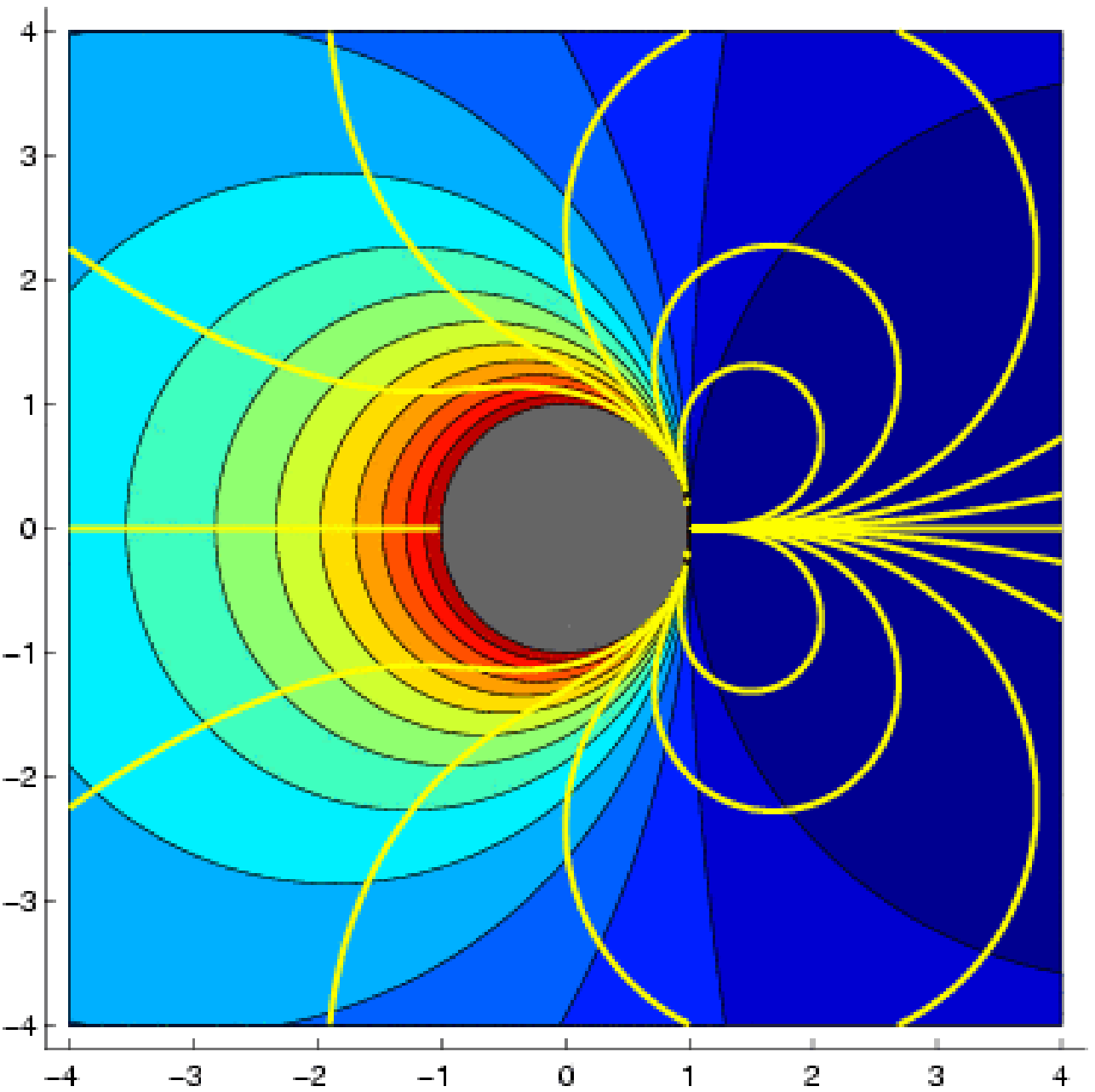}
\nopagebreak \caption{ Concentration profiles (contour plots) and
potential-flow streamlines (yellow) for steady, linear advection-diffusion
layers around various absorbing surfaces (gray) at $\Pe=1$. All solutions
are given by Eq.~(\protect\ref{eq:difflayer}), where $w=f(z)$ is a
conformal map to the upper half plane (top left). The color scale applies
to all panels in Figs.~\protect{\ref{fig:cd}} and \protect\ref{fig:circle}.
\label{fig:cd} }
\end{figure}

Let us rederive a classical solution in the upper half plane, $w = \xi +
i\eta$ ($\eta>0$), which we will then map to other geometries. As
shown in the top left panel of Fig.~\ref{fig:cd}, consider a straining
velocity field, $\phi = \Real \Phi$, $\Phi = w^2$, $u =
\overline{\Phi^\prime} = 2 \overline{w} = 2\xi - 2i\eta$, which
advects a concentrated fluid, $c(\xi,\infty)=1$, toward an absorbing
wall on the real axis, $c(\xi,0)=0$.  Since the $\eta$-component of
the velocity (toward the wall) is independent of $\xi$, as are the
boundary conditions, the concentration depends only on $\eta$. The
scaling function, $c(\xi,\eta; \Pe) = S(\sqrt{\Pe} \eta) =
S(\tilde{\eta})$, satisfies
\begin{equation}
-2\tilde{\eta} \, S^\prime = (b(S)\, S^\prime)^\prime , \ \ \ S(0)=0, \ \ \
S(\infty)=1 , \label{eq:bv}
\end{equation}
which is straightforward to solve, at least numerically.
For $b(S)=1$, Equation (\ref{eq:bv}) has a simple, analytical solution,
$S(\tilde{\eta}) = \erf (\tilde{\eta})$ (e.g. Cummings {\it et al.} 1999).

If extended to the entire $w$-plane, where two fluids of different
concentrations flow towards each other, this solution also describes a
Burgers' vortex sheet under uniform strain (Burgers 1948).  In that
case, $(\phi_\xi,\phi_\eta,c)$ is a three-dimensional velocity field
satisfying the Navier-Stokes equations, and $\Pe$ is the Reynolds
number. Inserting a boundary, such as the stationary wall on the real
axis, however, is not consistent with Burgers' solution because the
no-slip condition cannot be satisfied. The wall is crucial for
conformal mapping to other geometries because it enables
singularities to be placed in the lower half plane.

For every conformal map to the upper half plane, $w = f(z)$, we obtain
a solution,
\begin{equation}
\phi = \Real f(z)^2 \ \ \ \mbox{and } \ \ \ c = S\left( \sqrt{\Pe} \Imag
f(z) \right)  \ \ \ \mbox{ for } \Imag\, f(z) \geq 0 \label{eq:difflayer}
\end{equation}
which describes the nonlinear advection-diffusion layer in a potential flow
of concentrated fluid around the leading edge of an absorbing object.
For a linear diffusivity, $S(\tilde{\eta}) = \erf \tilde{\eta}$,
various examples are shown in Fig.~\ref{fig:cd}. The choice, $f(z) =
\sqrt{z}-a$, in the middle left panel, describes a parabolic leading
edge, $x = (y/2\alpha)^2-\alpha^2$, where $\alpha = \Imag a \geq
0$. The limit of uniform flow past a half plate ($a=0$), in the upper
right panel is a special case discussed below.

Another classical map, $f(z) = z^{\pi/(2\pi-\beta)}$, describes a wedge
of opening angle, $\beta$, as shown in the middle right panel for
$\beta=\pi/2$ (after a rotation by $\pi/4$). The half plate ($\beta=0$) and
the flat wall ($\beta=\pi$) discussed above are special cases. The
diffusive flux on the surface from Eq.~(\ref{eq:nflux}), $|\nabla \phi|
\propto \sqrt{\Pe}\, r^{-\nu}$, is singular for acute angles, $\beta<\pi$.
The geometry-dependent exponent, $\nu = (\pi-\beta)/(2\pi-\beta)$, is the
same for pure diffusion to the wedge, $\phi_d \propto \Imag f(z)$
(Barenblatt 1995). This insensitivity to $\Pe$ is a signature of the
Equivalence Theorem, as explained below.

\begin{figure}
\includegraphics[width=2.5in]{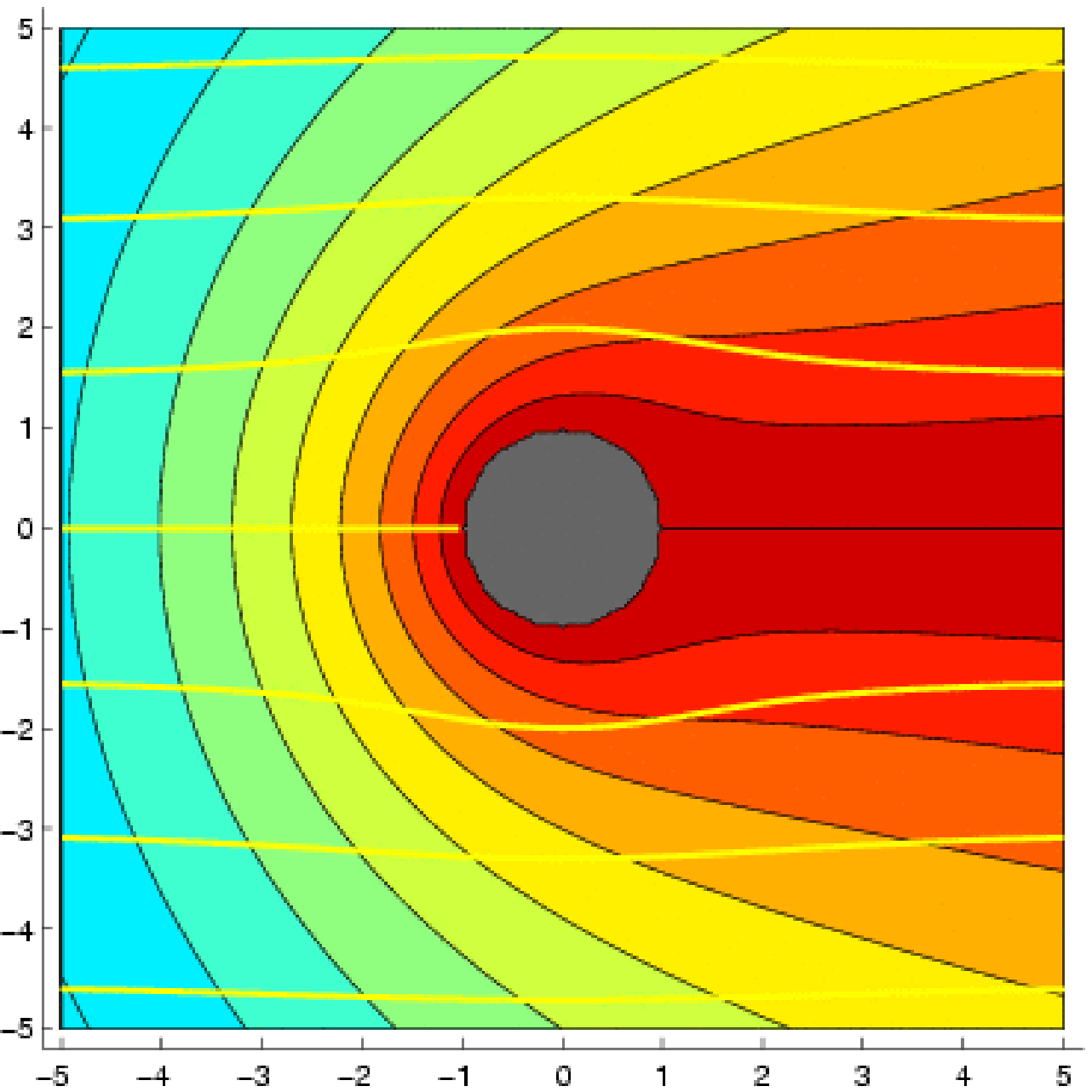} \nolinebreak
\includegraphics[width=2.5in]{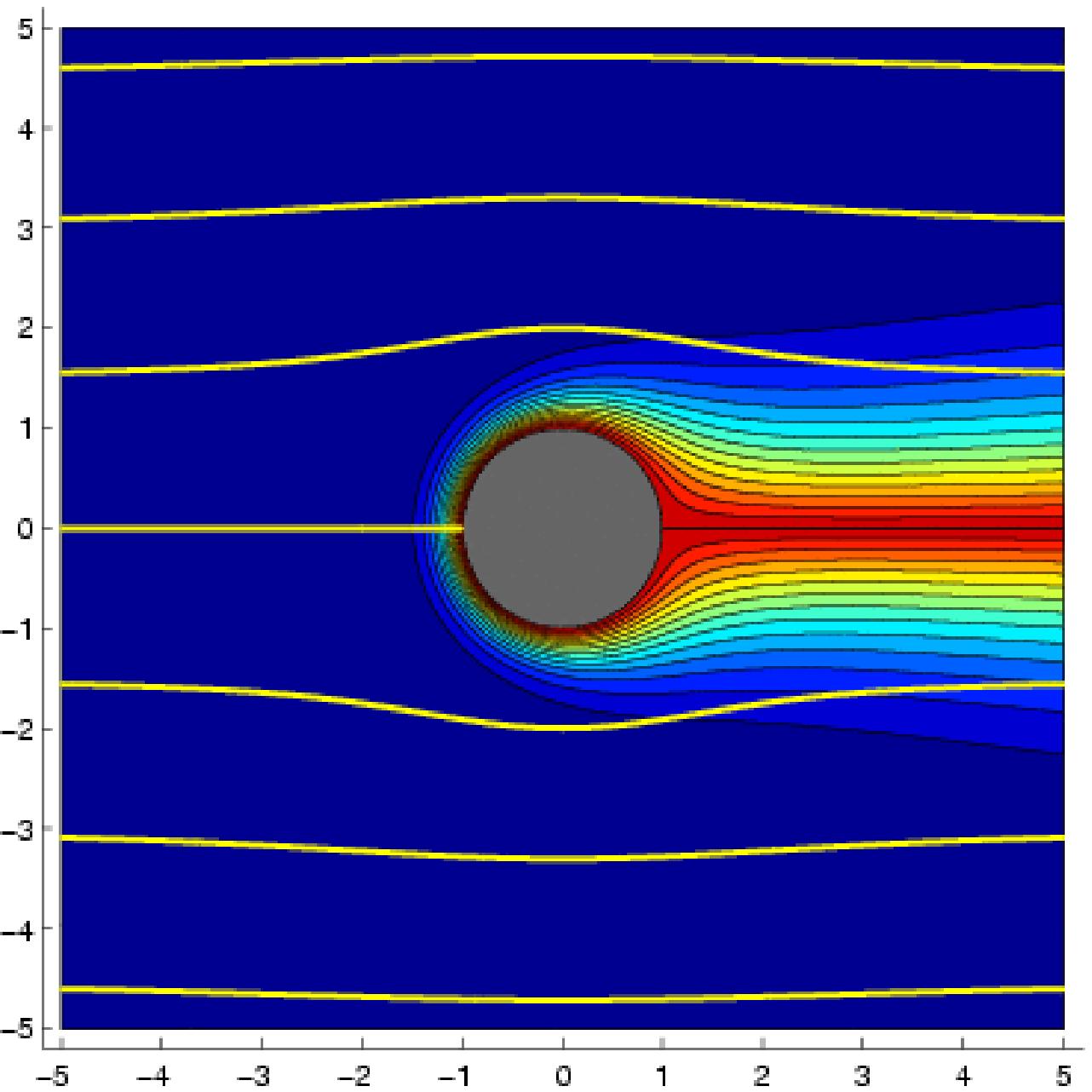}
\caption{ The steady linear advection-diffusion layer around a cylindrical
rim on a flat plate at $\Pe=0.1$ (left) and $\Pe=10$ (right).
\label{fig:circle} }
\end{figure}

The less familiar mapping, $f(z) = z^{1/2} + z^{-1/2}$, which plays a
crucial role in non-Laplacian growth problems (see below), places a
cylindrical rim on the end of a semi-infinite flat plate, as shown in the
lower left panel. The solution has a pleasing form in polar coordinates,
\begin{eqnarray}
\phi &=& \left(r + \frac{1}{r}\right) \cos\theta \label{eq:phirim} \\ c &=&
\erf\left[ \sqrt{\Pe} \left(\sqrt{r} -
\frac{1}{\sqrt{r}}\right)\sin\frac{\theta}{2}\right] \label{eq:crim}
\end{eqnarray}
where we have shifted the velocity potential, $\Phi = f(z)^2-2 = z +
z^{-1}$.  Far from the rim, we recover the half-plate similarity
solution, since $f(z)\sim\sqrt{z}$ as $|z|\rightarrow\infty$, but
close to the rim, as shown in Fig.~\ref{fig:circle}, there is a
nontrivial dependence on $\Pe$. For $\Pe\gg 1$, a boundary layer of
$O(\Pe^{-1/2})$ thickness forms on the front of the rim and extends to
within an $O(\Pe^{-1/2})$ distance from the rear stagnation point.

The flux density is easily calculated in the $w$-plane and then mapped to
the $z$-plane using Eq.~(\ref{eq:Ftrans}):
\begin{equation}
F_z\, = \, 2\, \overline{f^\prime(z) f(z)}\, \Pe\, S\left(\sqrt{\Pe} \Imag
f(z)\right) -\overline{f^\prime(z)} \, \sqrt{\Pe}\,
S^\prime\left(\sqrt{\Pe} \Imag f(z)\right) \label{eq:Fedge}
\end{equation}
where the first term describes advection and the second, diffusion. The
lines of advective and diffusive flux, which are level curves of $\Imag
f(z)^2$ and $\Real f(z)$, respectively, are independent of $\Pe$ and
$b(c)$, as required by the Equivalence Theorem. In particular, the
diffusive flux lines have the same shape for any flow speed or nonlinear
diffusivity as in the case of simple linear diffusion ($\Pe = 0$, $b(c) =
1$, $c \propto \Imag f(z)$), even though advection and nonlinearity
affect the lines of total flux.

The lines of total flux, called `heatlines' in thermal
advection-diffusion, are level curves of the `heat function'
\footnote{ Sen \& Yang (2000) have recently shown that the heat
function satisfies Laplace's equation, $\tilde{\del}^2 H = 0$, in certain
potential-dependent coordinates, $\tilde{\del}\equiv \re^{-\Pe\, \phi}\,
\del $.  This might seem related to our theorems, but it does not provide a
basis for conformal mapping of the domain because the coordinate
transformation is not analytic. Its value is also limited by the fact that
the boundary conditions on $H$ are not known {\it a priori}. For example,
on a surface where the concentration is specified, the unknown flux is
also required. These difficulties underscore the fact that the solutions
of Eq.~(\ref{eq:cd}) are fundamentally non-harmonic.  } (Kimura \& Bejan
1983), which we define in complex notation via $\nabla H = i F $.  For a
linear diffusivity,
we integrate Eq.~(\ref{eq:Fedge}) to obtain the heat function for any
conformal mapping,
\begin{equation}
H = 2 \,\Real f(z) \left[ \Pe\, (\Imag f(z)) \, \erf\left(\sqrt{\Pe}
\Imag f(z)\right)\, + \,
\sqrt{\frac{\Pe}{\pi}}\, \exp\left(-\Pe\, (\Imag f(z))^2 \right)
\right] , \label{eq:H}
\end{equation}
which shows how the total-flux lines cross over smoothly from fluid
streamlines outside the diffusion layer ($H \sim \Pe \Imag f(z)^2$, $\Pe
\Imag f(z) \gg 1$) to diffusive-flux lines near the absorbing surface ($H
\sim 2 \sqrt{\Pe/\pi} \, \Real f(z)$, $\Pe \Imag f(z) \ll 1$).

On the absorbing surface, $\Imag f(z) = 0$, the flux density is purely
diffusive and in the normal direction. Its spatial distribution is
determined {\it geometrically} by the conformal map,
\begin{equation}
|F_z|\, = \, \sqrt{\Pe} \, S^\prime(0) \, |f^\prime(z)| \ \ \ \mbox{
on } \ \Imag f(z) = 0 ,  \label{eq:nflux}
\end{equation}
and only its magnitude depends on $\Pe$, as predicted by the Equivalence
Theorem. (For a linear diffusivity, $S^\prime(0) = 2/\sqrt{\pi}$.)  What
appears to be the only previous result of this kind is due to Koplik {\it
et al.} (1994, 1995) in the context of tracer dispersion by linear
advection-diffusion in porous media. In the case of planar potential flow
from a dipole source to an equipotential absorbing sink, they proved that
the spatial distribution of surface flux is independent of $\Pe$. Here we
see that the same conclusion holds for all similarity solutions to
Eq.~(\ref{eq:cd}), even if (i) diffusive flux is not directed along
streamlines; (ii) the diffusivity is a nonlinear function of the
concentration; and (iii) the domain is on a curved surface.

\subsection{ Streamline Coordinates }
In proving their equivalence theorem, Koplik {\it et al.} (1994, 1995)
transform Eq.~ (\ref{eq:cd}) in the linear case, $b(c)=1$, to
`streamline coordinates',
\begin{equation}
\Pe c_\phi = c_{\phi\phi} + c_{\psi\psi} ,  \label{eq:stream}
\end{equation}
where
$\Phi = \phi + i \psi$
is the complex potential, $\phi$, the velocity potential, and $\psi$,
the streamfunction.  Because the independent and dependent variables
are interchanged, this is a type of hodographic transformation
(Whitham 1974; Ben Amar \& Poir\'e 1999). The physical interpretation
of Eq.~(\ref{eq:stream}) is that advection (the left-hand side) is
directed along streamlines, while diffusion (the right-hand side) is
also perpendicular to the streamlines, along iso-potential lines.  In
high-Reynolds-number fluid mechanics, this is a well known trick due
to Boussinesq (1905) still in use today (Hunt \& Eames
2002). Streamline coordinates are also used in Maksimov's method for
dendritic solidification from a flowing melt (Cummings {\it et al.}
1999).

Boussinesq's transformation is simply a conformal mapping to a
geometry of uniform flow. Any obstacles in the flow are mapped to line
segments (branch cuts of the inverse map) parallel to the
streamlines. Among the solutions (\ref{eq:difflayer}), streamline
coordinates correspond to the map, $f(z) = \sqrt{z}$, from a plane of
uniform flow past an absorbing flat plate on the positive real axis
(the branch cut), as shown in the top right panel of
Fig.~\ref{fig:cd}. In this geometry, we have the boundary-value
problem, $\Pe\, \frac{\partial c}{\partial x} = \del^2 c$, $c(x>0,0) =
0$, $c(-\infty,y)=1$, which Carrier {\it et al.} (1983) have solved
using the Weiner-Hopf technique. More simply, Greenspan has introduced
parabolic coordinates (as in Greenspan 1961), to immediately obtain
the similarity solution derived above, $c(x,y) = \erf(\sqrt{\Pe}\,
\eta)$, where $2\eta^2 = -x + \sqrt{x^2 + y^2}$.  The reason why this
solution exists, however, only becomes clear after conformal mapping
to {\it non-streamline coordinates} in the upper half plane. (See also
Cummings {\it et al.} 1999.)

As this example illustrates, streamline coordinates are not always 
convenient, so it is useful to exploit the possibility of conformal
mapping to other geometries.  For similarity solutions, it is easier
to work in a plane where the diffusive flux lines are parallel.
Streamline coordinates are also often poorly suited for numerical
methods because stagnation points are associated with branch-point
singularities. This is especially problematic for free boundary
problems: For flows toward infinite dendrites, it is easier to
determine the evolving map from a half plane (Cummings {\it et al.}
1999); for flows past finite growing objects, it is easier to map from
the exterior of the unit circle (Bazant, Choi \& Davidovitch 2003).

\subsection{ Non-similarity Solutions for Finite Absorbing Objects }
\label{sec:finite}

It is tempting to try to eliminate the plate from the cylindrical rim
in Fig.~\ref{fig:circle} by conformal mapping from the exterior of a
finite object to the upper half plane. Any such mapping in
Eq.~(\ref{eq:difflayer}), however, requires a quadrupole point source
of flow (mapped to $\infty$) on the object's surface. This is
illustrated in the lower right panel of Fig.~\ref{fig:cd} by a
M\"obius transformation from the exterior of the unit circle, $f(z) =
(1+z)/i(1-z)$, where a source at $z=1$ ejects concentrated
fluid in the $+1$ direction and sucks in fluid along the $\pm i$
directions.
Thus we see that, due to the boundary conditions at $\infty$, uniform
flow past an absorbing cylinder (or any other finite object) is in a
different class of solutions, where the diffusive flux lines depend
nontrivially on $\Pe$. In streamline coordinates, this includes the
problem of uniform flow past a finite absorbing strip, which requires
solving Wijngaarden's integral equation (Cummings {\it et al.} 1999).

Here, we study only the high-$\Pe$ asymptotics of
advection-diffusion layers around finite absorbing objects.
Consider again the example of flow past a cylindrical rim on a flat
plate (Fig.~\ref{fig:circle}). Because disturbances in the
concentration decay exponentially upstream beyond an $O(\Pe^{-1/2})$
distance, removing the plate on the downstream side of the cylinder
has no effect in the limit $\Pe \rightarrow \infty$, except on the
plate itself (the branch cut), so the solution
~(\ref{eq:phirim})--(\ref{eq:crim}) is also asymptotically valid near
a finite absorbing cylinder (without the plate).

More generally, if $z=h(q)$ is  the conformal map from the exterior of
any singly connected finite object to the exterior of the unit circle,
then the non-harmonic concentration field has the asymptotic form,
\begin{equation}
c(q,\overline{q}) \sim \erf\left[ \sqrt{\Pe}\, \Imag\, \left( \sqrt{h(q)} +
\frac{1}{\sqrt{h(q)}} \right) \right] \label{eq:crimlim}
\end{equation}
as $\Pe \to\infty$ everywhere except in the wake near the pre-image of
the positive real axis, a branch cut corresponding to the `false
plate'. The convergence is not uniform, since the false plate always
spoils the approximation sufficiently far downstream, for a fixed $\Pe
\gg 1$.  The validity of Eq.~(\ref{eq:crimlim}) near the surface of
the object, however, allows us to calculate the normal flux density
using Eq.~(\ref{eq:nflux}),
\begin{equation}
\nhat\cdot \del c \sim  2 \sqrt{\frac{\Pe}{\pi}}
\sin\left(\frac{\theta}{2}\right)  \label{eq:rimflux}
\end{equation}
as $ \Pe \rightarrow\infty$ for all $\theta = \arg h(q) \gg
\Pe^{-1/2}$ away from the rear stagnation point, $\theta=0$.  The
limiting Nusselt number, $\Nu \sim 8\sqrt{\Pe/\pi}$, is also easily
calculated by mapping the rim (with the false plate) to the upper half
plane where the normal flux density is uniform, $2\sqrt{\Pe/\pi}$, on
a line segment of length four (from -2 to 2).

As explained in section 2(c), Equation~(\ref{eq:rimflux}) describes
the non-harmonic probability measure for fractal growth by steady
advection-diffusion in a uniform potential flow in the limit $\Pe
\rightarrow \infty$. This model, which we might call
`advection-diffusion-limited aggregation' (ADLA), is perhaps the
simplest generalization of the famous DLA model of Witten and Sander
(1981) allowing for more than one bulk transport process.  The
resulting competition between advection and diffusion produces a
crossover between two distinct statistical `phases' of growth.  As
expected from renormalization-group theory (Goldenfeld 1992), the
crossover connects `fixed points' of the growth measure, describing
self-similar dynamics. For small initial P\'eclet numbers, $\Pe(0)\ll
1$, the growth measure of ADLA is well approximated by the uniform
harmonic measure of DLA and the concentration by the similarity
solution, $c(q,\overline{q}) \propto \Imag \log h(q)$, but this is an
unstable fixed point.  
Regardless of the initial conditions, the P\'eclet number diverges,
$\Pe(t) = U\, L(t) / D \rightarrow \infty$, as the object grows, so
the concentration eventually approaches the new similarity solution in
Eq.~(\ref{eq:crimlim}). At this advection-dominated stable fixed
point, the growth measure obeys Eq.~(\ref{eq:rimflux}). The
$\sin\theta/2$ dependence causes anisotropic fractal growth at long
times favoring the direction of incoming, concentrated fluid, $\theta
= \pi$, and the total growth rate ($\Nu$) is proportional to
$\sqrt{\Pe(t)}$. Such analytical results serve to illustrate the power
of conformal mapping applied to systems of invariant equations.

\section{ Electrochemical Transport }
\label{sec:elec}

\subsection{ Simple Approximations and Conformal Mapping }

Conservation laws for gradient-driven fluxes also describe ionic
transport in dilute electrolytes. Because the complete set of
equations and boundary conditions (below) are nonlinear and rather
complicated, the classical theory of electrochemical systems involves
a hierarchy of approximations (Newman 1991). Conformal mapping has
long been applied in the simplest case where the current density,
$\Jb$, is proportional to the gradient of a harmonic function, $\phi$,
the electrostatic potential (Moulton 1905; Hine 1956).

This approximation, the `primary current distribution', describes the
linear response of a homogeneous electrolyte to a small applied
voltage or current, as well as more general conduction in a supporting
electrolyte (a great excess of inactive ions). The assumptions of
Ohm's Law, $\Jb = \sigma \Eb = - \sigma \del\phi$ (with a constant
conductivity, $\sigma$) and no bulk charge sources or sinks,
$\del\cdot\Jb=0$, are analogous to those of potential flow and
incompressibility describe above. Each electrode is assumed to be an
equipotential surface (see below), so the potential is simply that of
a capacitor --- harmonic with Dirichlet boundary
conditions. Naturally, classical conformal mapping from electrostatics
(Churchill \& Brown 1990; Needham 1997) have been routinely applied,
but it seems conformal mapping has never been applied to any more
realistic models of electrochemical systems.

The `secondary current distribution' introduces a kinetic boundary
condition, $\nhat\cdot\Jb = R(\phi)$, which equates the normal current
with a potential-dependent reaction rate, e.g. given by the
Butler-Volmer equation (see below). In this case, conformal mapping
could be of some use.  Although the boundary condition acquires a
non-constant coefficient, $|f^\prime|$, from Eq.~(\ref{eq:Ftrans}),
Laplace's equation is preserved.

A more serious complication in the `tertiary current distribution' is
to allow the bulk ionic concentrations to vary in space (but not
time). Ohm's law is then replaced by a nonlinear current-voltage
relation. Our main insight here is that conformal mapping can still be
applied in the usual way, even though the equations are nonlinear and
the potential, non-harmonic.

\subsection{ Dilute-Solution Theory }

In the usual case of a dilute electrolyte, the ionic concentrations,
$\{c_1, c_2, \ldots, c_N \}$, and the electrostatic potential, $\phi$,
satisfy the Nernst-Planck equations (Newman 1991), which have the form of
Eqs.~(\ref{eq:cons}) and (\ref{eq:Fi}), where the `advection' velocities,
$\ub_i = - z_i e \mu_i \del \phi$, are due to migration in the electric
field, $\Eb = -\del \phi$. Here, $z_i e$ is the charge (positive or
negative) and $\mu_i$ the mobility of the $i$th ionic species. The
diffusivities are given by the Einstein relation, $D_i = k_B T \mu_i$,
where $k_B$ is Boltzmann's constant and $T$, the temperature.
Scaling concentrations to a reference value, $C$, potential to the thermal
voltage, $kT/e$, length to a typical electrode separation, $L$, and
assuming that $D_i$, $T$, and $\varepsilon$ are constants, the steady-state
equations take the dimensionless form,
\begin{equation}
\del^2 c_i + z_i \del\cdot(c_i\del\phi) =
0 . \label{eq:ion}
\end{equation}
The ionic flux densities are $ \Fb_i = -\del c_i
-z_i c_i \del \phi$ (scaled to
$D_i C/L$).

Because dissolved ions are very effective at charge screening,
significant diffuse charge can only exist in very thin ($1-100$nm)
interfacial double layers, where boundary conditions break the
symmetry between opposite charge carriers. The `bulk' potential
(outside the double layers) is then determined implicitly by the
condition of electroneutrality
(Newman 1991), $\sum_{i=1}^N z_i c_i = 0$, which is trivially
conformally invariant. Therefore, the most common model of steady
electrochemical transport, Eq.~(\ref{eq:ion}), satisfies the
assumptions of the Conformal Mapping Theorem for any number of ionic
species ($N \geq 2$).  Although the equations differ from those of
advection-diffusion in a potential flow, we can still map to {\it
electric-field coordinates} (the analog of streamline coordinates), or
any other convenient geometry.

Although the equations are conformally invariant, the boundary
conditions are so only in certain limits. General boundary conditions
express mass conservation, either $\nhat\cdot\Fb_i=0$ for an inert
species, or 
\begin{equation}
\nhat \cdot \del \Fb_i = R_i(c_i,\phi),
\label{eq:fluxbc}
\end{equation}
for an active species at an electrode, where $R_i(c_i,\phi)$ is the
Faradaic reaction-rate density (scaled to $D_iC/L$). It is common to
assume Ahrrenius kinetics,
\begin{equation}
R_i(c_i,\phi) = k_+ c_i e^{z_i \alpha_+(\phi-\phi_e)} - k_- c_r e^{-z_i
\alpha_- (\phi-\phi_e)}  \label{eq:R}
\end{equation}
where $k_+$ and $k_-$ are rate constants for deposition and
dissolution, respectively (scaled to $D_i/L$), $\alpha_\pm$ are
transfer coefficients, $c_r$ is the concentration of the reduced
species (scaled to $C$) and $\phi_e$ is the electrode potential
(scaled to $kT/e$). Taking diffuse interfacial charge into account
somewhat modifies $R(c_i,\phi)$, but the basic structure of
Eq.~(\ref{eq:fluxbc}) is unchanged (Newman 1991; Bonnefont {\it et
al.} 2001). Conformal mapping introduces a non-constant coefficient,
$|f^\prime|$, in Eq.~(\ref{eq:fluxbc}), but conformal invariance is
restored in the case of `fast reactions' ($k_+\gg 1, k_- c_r \gg 1$),
in which equilibrium conditions prevail, $R=0$, even during the
passage of current. For a single active species (say $i=1$), the bulk
potential at an electrode is then given by the (dimensionless) Nernst
equation,
\begin{equation}
\phi-\phi_e = \Delta\phi_{eq} = - \frac{\log(k c_1)}{z_1 (\alpha_+ +
\alpha_-)} , \label{eq:nernst}
\end{equation}
where $k = k_+/k_- c_r$ is an equilibrium constant\footnote{ Expressing
Eq.~(\ref{eq:R}) in terms of the `surface overpotential', $\eta_s = \phi -
\phi_e - \Delta\phi_{eq}$, yields the more familiar Butler-Volmer equation
(Newman 1991).}.

\begin{figure}
\begin{center}
\includegraphics[width=3in]{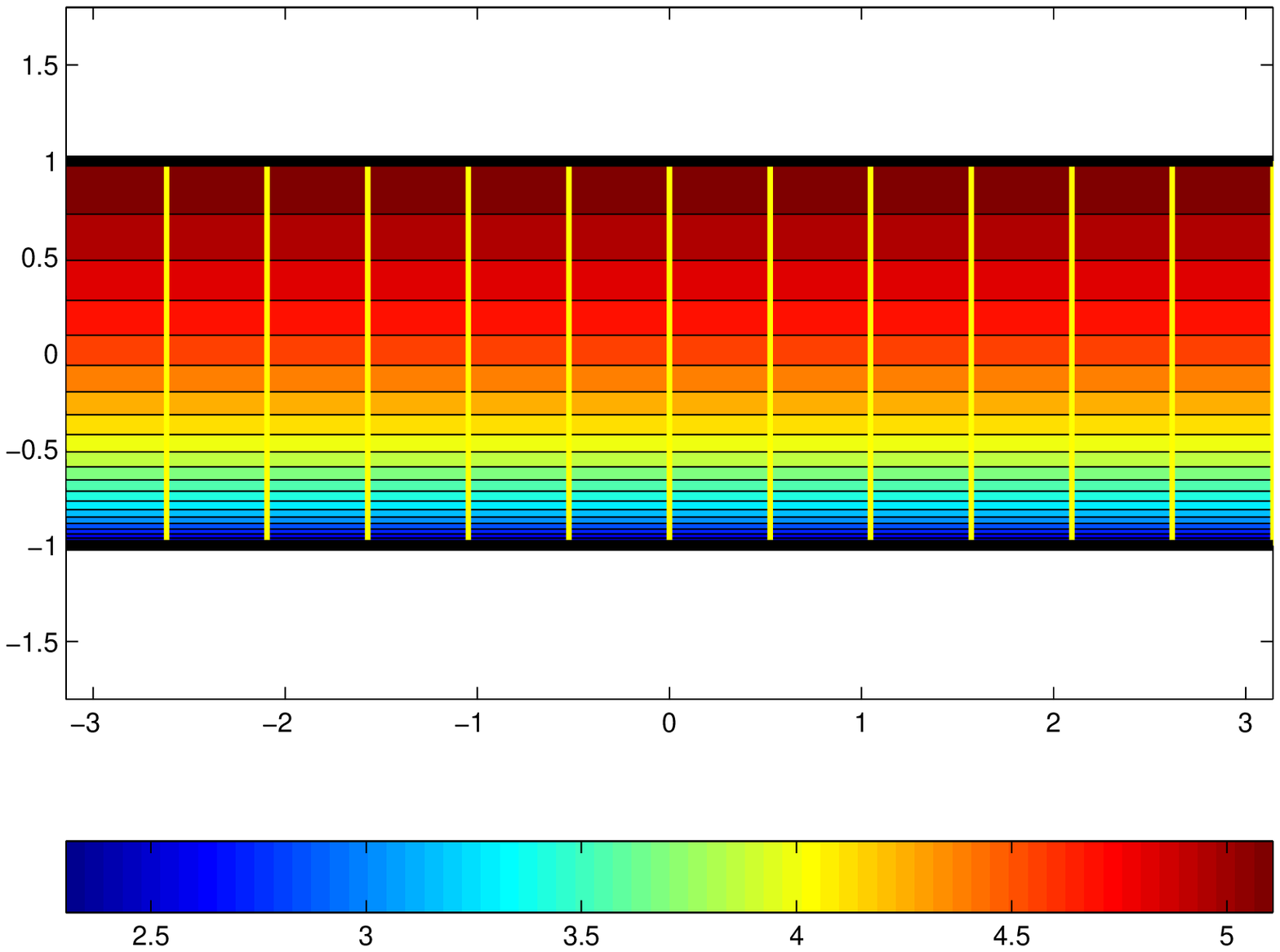}\vspace{0.1in}\\
\includegraphics[width=2.6in]{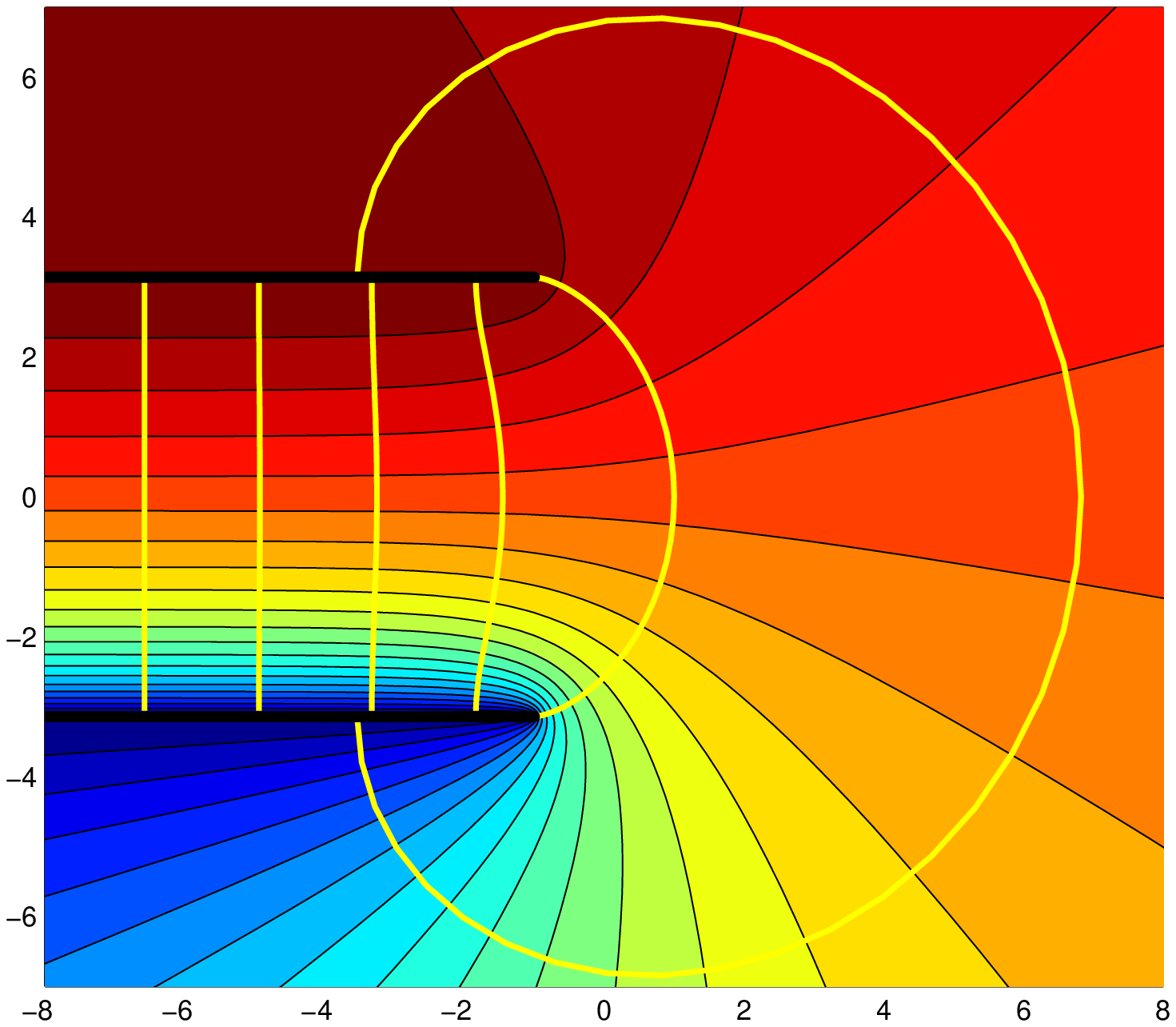} \nolinebreak
\includegraphics[width=2.3in]{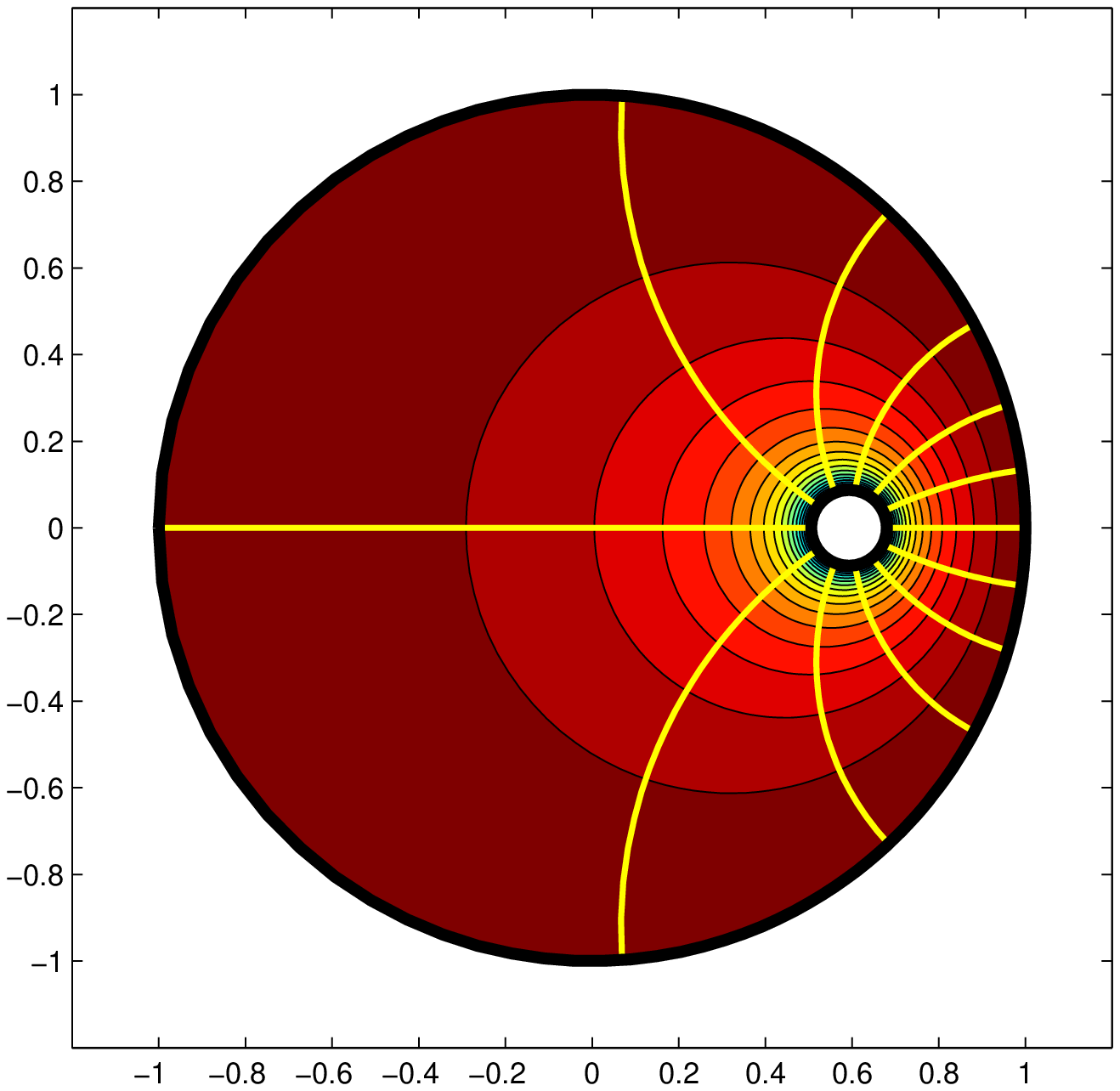}
\end{center}
\caption{ Similarity solutions for the electrostatic potential
(contour plot) and current/electric-field lines (yellow) in a binary
electrolyte at 90\% of the limiting current ($k=1$). The simple
solution for parallel-plate electrodes (top) is conformally mapped to
semi-infinite plates (bottom left) and misaligned coaxial cylinders
(bottom right). \label{fig:fringe} }
\end{figure}

\subsection{ Conformal Mapping with Concentration Polarization }

The voltage across an electrochemical cell is conceptually divided
into three parts (Newman 1991): (i) the `Ohmic polarization' of the
primary current distribution, (ii) the `surface polarization' of the
secondary current distribution, and (iii) `concentration
polarization', the remaining voltage attributed to non-uniform bulk
concentrations. Although concentration polarization can be
significant, especially at large currents in binary electrolytes, it
is difficult to calculate. Analytical results are available only for
very simple geometries (mainly in one dimension), so our method easily
produces new results.

For example, consider a symmetric binary electrolyte ($N=2$) of charge
number, $z = z_+ = - z_-$, where the concentration, $c = c_+ = c_-$, and
the potential satisfy,
\begin{equation}
\del^2 c = 0 \ \ \ \mbox{ and } \ \ \ \del\cdot(c\del\phi) = 0 .
\end{equation}
(The concentrations are harmonic only for $N=2$.) Assuming that anions are
chemically inert yields an invariant zero-flux condition at each electrode,
$\nhat\cdot(c\del\phi - \del c)=0 $,
and (to break degeneracy) a constraint on the integral of $c$, which sets
the total number of anions (Bonnefont {\it et al.} 2001). In the limit of
fast reactions, the bulk potential at each electrode is given by the Nernst
equation, $\phi = \phi_e - \log k c$, where we scale $\phi$ to $k_BT/ze$
and assume $\alpha_+ - \alpha_- = 1$.

A class of similarity solutions is obtained by conformal mapping, $w =
f(z)$, to a strip, $-1 < \Imag w < 1$, representing parallel-plate
electrodes.  We set $\phi_e=0$ at the cathode ($\Imag w = -1$) and
$\phi_e = V$, the applied voltage (in units of $k_BT/ze$), at the
anode ($\Imag w =1$). We then solve $c^{\prime\prime}=0$ and
$(c\phi^\prime)^\prime=0$ with appropriate boundary conditions to
obtain a general solution for any conformal mapping to the strip:
\begin{equation}
c = 1 + J\, \Imag f(z), \ \ \ \phi = \log\left( \frac{ 1 + J\, \Imag f(z)
}{k (1-J)^2} \right),
\end{equation}
where $J = \tanh(V/4)$ is the uniform current density in the strip,
scaled to its limiting value, $J_{lim} = 2zeD_+C/L$. As $J \rightarrow
1$, strong concentration polarization develops near the cathode, as
shown in Fig.~\ref{fig:fringe} for $J=0.9$.  At $J=1$, the bulk
concentration at the cathode vanishes, and the cell voltage diverges
due to diffusion limitation.

The classical conformal map, $z = f^{-1}(w) = \pi w + e^{\pi w}$
(Churchill \& Brown 1990), unfolds the strip like a `fan' to cover the
$z$-plane and maps the electrodes onto two half plates ($\Imag z = \pm
\pi$ , $\Real z < -1$). As shown in Fig.~\ref{fig:fringe}, this
solution describes the fringe fields of semi-infinite, parallel-plate
electrodes.  The field and current lines are cycloids, $z_a(\eta) =
\pi a+i\pi\eta + e^{\pi a} e^{i\pi\eta}$, as in the limit of a
harmonic potential at low currents, $\phi \sim J \, \Imag f(z) - \log
k$. At high currents, the {\it magnitude} of the electric field is
greatly amplified near the cathode (the lower plate) by concentration
polarization, but the {\it shape} of the field lines is always the
same. This conclusion also holds for all other conformal mappings to
the strip, such as the M\"obius-log transformation, $w = f(z) = i( 1 +
\log(5z-3)/(5-3z))$, in Fig.~\ref{fig:fringe} from the region between
two non-concentric circles.

It is interesting to note that the Equivalence Theorem applies to some
physical situations and not others.  Similarity solutions like the
ones above can only be derived for {\it two} equipotential electrodes
by conformal mapping to a strip, where the current is uniform. In all
such geometries, the electric field lines have the same shape as in
the primary current distribution. For {\it three or more}
equipotential electrodes, however, this is no longer true because
conformal mapping to the strip is topologically impossible, and thus
similarity solutions do not exist. When the bulk potential varies at
the electrodes according to Eq.~(\ref{eq:fluxbc}), the electric field
lines generally differ from both the primary and secondary current
distributions, even for just two electrodes.

\section{ Conclusion }
\label{sec:concl}

We have observed that the nonlinear system of equations
(\ref{eq:geneq}) involving `dot products of two gradients' is
conformally invariant. This has allowed us to extend the classical
technique of conformal mapping to some non-harmonic functions arising
in physics. Examples from transport theory are steady conservation
laws for gradient-driven fluxes, Eq.~(\ref{eq:Fi}).  For one variable,
the equations in our class (including some familiar examples in
nonlinear diffusion) can always be reduced to Laplace's equation.  For
two or more variables, the general solutions are not simply related to
harmonic functions, but all similarity solutions exhibit an
interesting geometrical equivalence.


For two variables, there is one example in our class, steady
advection-diffusion in a potential flow, to which conformal mapping
has previously been applied.  In this case, our method is equivalent
to Boussinesq's streamline coordinates, but somewhat more general. A
nonlinear diffusivity is also allowed, and the mapping need not be to
a plane of uniform flow (parallel streamlines). In a series of
examples, we have considered flows past absorbing leading edges and
have generalized a recent equivalence theorem of Koplik, Redner, and
Hinch (1994, 1995). We have also considered the flows past finite
absorbing objects at high P\'eclet number.

Our class also contains the Nernst-Planck equations for steady, bulk
electrochemical transport, for which very few exact solutions are
known in more than one dimension. In electrochemistry, conformal
mapping has been applied only to harmonic functions, so we have
presented some new results, such as the concentration polarizations
for semi-infinite, parallel-plate electrodes and for misaligned
coaxial electrodes. More generally, we have shown that Ohm's Law gives
the correct spatial distribution (but not the correct magnitude) of
the electric field on any pair of equipotential electrodes in two
dimensions, even if the transport is nonlinear and non-Laplacian,
although this is not true for three or more electrodes. Such results
could be useful in modeling micro-electrochemical systems, where
steady states are easily attained (due to short diffusion lengths) and
quasi-planar geometries are often arise.

As mentioned thoughout the paper, our results can be applied to a
broad class of moving free boundary problems for systems of
non-Laplacian transport equations (Bazant, Choi \& Davidovitch
2003). In contrast, the vast literature on conformal-map dynamics
(cited in section 1) relies on complex-potential theory, which only
applies to Laplacian transport processes.  Nevertheless, standard
formulations, such as the Polubarinova-Galin equation for continuous
Laplacian growth (Howison 1992) and the Hastings-Levitov (1998) method
of iterated maps for DLA, can be easily generalized for coupled
non-Laplacian transport processes in our class. In the stochastic
case, non-harmonic probability measures for fractal growth can be
defined on any convenient contour, such as the unit circle. As an
example, we have derived the stable fixed point of the growth measure
for an arbitrary absorbing object in a uniform background potential
flow, Eq.~(\ref{eq:rimflux}). This sets the stage for
conformal-mapping simulations of ADLA, which might otherwise seem
intractable.

\bigskip
\bigskip
\noindent The author is grateful to J. Choi for the color figures;
A. Ajdari, M. Ben Amar, J. Choi, K. Chu, D. Crowdy, B. Davidovitch,
I. Eames,
E. J. Hinch, H. K. Moffatt, and A. Toomre for helpful comments on the
manuscript; MIT for a junior faculty leave; and ESPCI for hospitality
and support through the Paris Sciences Chair. This work was also
supported by the MRSEC Program of the National Science Foundation
under award number DMR 02-13282. \nopagebreak

\end{document}